\documentclass{mn2e}
\usepackage{epsfig}
\usepackage{times}
\begin{document}

\title[XTE J1701-462] {The variable radio counterpart and possible
large-scale jet of the new Z-source XTE J1701-462} 
\author[Fender et al.]  {R. P. Fender$^1$\thanks{email:rpf@phys.soton.ac.uk},
M. Dahlem$^2$, J. Homan$^3$, S. Corbel$^4$, R. Sault$^2$,
T.M.Belloni$^5$\\ $^1$ School of Physics and Astronomy, University of
Southampton, Highfield, Southampton, SO17 1BJ, UK\\ $^2$ CSIRO/ATNF,
Paul Wild Observatory, Locked Bag 194, Narrabri NSW 2390, Australia \\
$^3$ Kavli Institute for Astrophysics and Space Research,
Massachusetts Institute of Technology, 77 Massachusetts Avenue,
Cambridge, MA 02139, USA\\ $^4$ AIM-Unit\'e Mixte de Recherche
CEA-CNRS, Universi\'e Paris VII, UMR 7158, CEA Saclay, Service
d'Astrophysique, F-91191 Gif-sur-Yvette, France\\ $^5$ INAF -
Osservatorio Astronomico di Brera, via Bianchi 46, 23807 Merate, Italy
\\} \maketitle

\begin{abstract}
We report radio observations, made with the Australia Telescope
Compact Array, of the X-ray transient XTE J1701-462. This system has
been classified as a new `Z' source, displaying characteristic
patterns of behaviour probably associated with accretion onto a low
magnetic field neutron star at close to the Eddington limit. The radio
counterpart is highly variable, and was detected in six of sixteen
observations over the period 2006 January -- April. The coupling of
radio emission to X-ray state, despite limited sampling, appears to be
similar to that of other `Z' sources, in that there is no radio
emission on the flaring branch. The mean radio and X-ray luminosities
are consistent with the other Z sources for a distance of 5--15 kpc.
The radio spectrum is unusually flat, or even inverted, in contrast to
the related sources, Sco X-1 and Cir X-1, which usually display an
optically thin radio spectrum. Deep wide-field observations indicate
an extended structure three arcminutes to the south which is aligned
with the X-ray binary. This seems to represent a significant
overdensity of radio sources for the field and so, although a
background source remains a strong possibility, we consider it
plausible that this is a large-scale jet associated with XTE
J1701-462.
\end{abstract}

\section{Introduction}

The relation between accretion and outflow is a key topic in modern
high energy astrophysics and offers us a unique opportunity to
understand the physics of distant, supermassive black holes in active
galactic nuclei (AGN) by studying nearby, rapidly varying objects such
as X-ray binaries. Much of the focus in recent years has been on black
holes and how accretion scales between those of mass $\sim
10$M$_{\odot}$ in XRBs and those of mass $\geq 10^5$M$_{\odot}$ in AGN
(Merloni, Heinz \& di Matteo 2003; Falcke, K\"ording \& Markoff 2004;
see also Maccarone, Gallo \& Fender (2004); K\"ording,
Corbel \& Falcke 2006; Wang, Wu \& Kong 2006; McHardy et al. 2006).

However, the neutron star XRBs represent an extremely valuable
`control sample'. These systems also produce dramatic jets
(e.g. Fomalont et al. 2001; Fender et al. 2004) over a wide range of
accretion rates (Migliari \& Fender 2005). By comparing the two
classes of object we can test the necessity of black hole-specific
physics, such as event horizons and static limits, on the observed
phenomena of accretion and jet production (e.g. K\"ording, Fender \&
Migliari 2006).

\begin{figure}
\epsfig{file=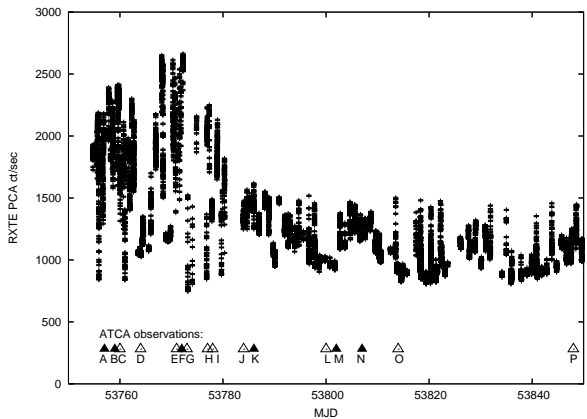, angle=270, width=8cm}
\caption{RXTE PCA monitoring of XTE J1701-462 over the period 2006
Jan-April, the first 100 days of the outburst. The epochs of the
sixteen ATCA radio observations in this period are indicated -- only
at epochs marked by filled symbols are there unambiguous radio
detections of the X-ray binary.}
\end{figure}

\begin{table*}
\begin{tabular}{ccccccccc}
Epoch & Detection & Start UT    & Stop UT     & Hr on target & S$_{4.8GHz}$      & S$_{8.6GHz}$ & X ? & X-ray state \\
A & yes    & Jan22 21:56 & Jan22 23:00 & 0.98         & $0.3 \pm 0.15$ & $0.4 \pm 0.1$ & no & HB/NB \\
B & yes     & Jan24 22:25 & Jan25 05:40 & 6.32         & $0.37 \pm 0.06$ & $0.36 \pm 0.05$ & partial & upper $\rightarrow$ lower HB\\
C & poss       & Jan25 23:15 & Jan26 00:54 & 1.49         & $<0.18$ & $<0.27$ & yes & FB \\
D & poss      & Jan29 04:18 & Jan29 05:23 & 0.93         & $<0.24$ & $<0.42$ & no & upper HB \\
E & no    & Feb05 03:20 & Feb05 05:02 & 1.34         & $<0.24$ & $<0.33$ & no & mid HB \\ 
F & yes       & Feb06 18:37 & Feb06 21:35 & 2.58         & $0.67 \pm 0.09$ & $0.55 \pm 0.07$ & no & NB / FB \\
G & no    & Feb07 20:10 & Feb07 21:04 & 0.75         & $<3.5$ & $<1.0$ & no & FB\\
H & no      & Feb11 01:37 & Feb11 01:57 & 0.34         & $<0.42$ & $<0.8$ & no & mid HB \\
I & no     & Feb12 00:45 & Feb12 01:49 & 0.98         & $<1.8$ & $<0.9$ & no & upper HB \\
J & no      & Feb18 19:31 & Feb18 21:00 & 1.19         & $<0.15$ & $<0.3$ & no & mid / lower NB\\
K & yes     & Feb20 18:40 & Feb20 21:06 & 1.91         & $0.2 \pm 0.1$ & $0.35 \pm 0.1$ & no & mid NB \\

L & no      & Mar06 19:23 & Mar06 21:05 & 1.33         & $<0.27$ & $<0.27$ & no & NB / FB vertex \\
M & yes      & Mar08 16:30 & Mar08 21:00 & 3.96         & $1.26 \pm 0.05$ & $1.60 \pm 0.07$ & yes & NB / HB \\
N & yes    & Mar13 15:19 & Mar13 22:53 & 6.64         & $0.32 \pm 0.03$ & $0.35 \pm 0.06$ & no & HB \\
O & poss     & Mar20 12:10 & Mar20 22:00 & 8.35         & $<0.15$ & $<0.15$ & no & mid / lower NB\\
P & no & Apr 23 14:51 & Apr 23 20:39 & 5.02 & $<0.05$ &$<0.1$ & no[?] & FB\\
\end{tabular}
\caption{Summary of radio observations of XTE J1701-462. Column 1
gives the Epoch label, which is used in the discussion in the main
text, and also Fig 1. Column 2 indicates whether or not there was a
detection of the radio source; only for those unambiguous detections,
where yes is indicated, do we provide flux densities. Columns 3-5 give
start and stop times and total hours on-source. Columns 6 and 7 give
detections or upper limits at 4.8 and 8.6 GHz respectively (note that
in some cases we have indicated detections which appear to be less
than three times the error -- this is due to uncertainties in the flux
calibration for short observations, and does not indicate the
reliability of the detection; this particularly applies to epochs A
and K). All measurements are based upon naturally-weighted maps, and
all upper limits are 3 times the r.m.s. noise in the area of the
source. Column 8 indicates whether or not the observation was strictly
simultaneous with an observation by RXTE, and column 9 indicates our
best estimate of the branch or branches of the Z occupied by the
source at, or close to, the time of the radio observation.}
\end{table*}

The 'Z sources' are the six (or seven, if you include Cir X-1) most
luminous neutron star X-ray binaries in our galaxy, persistently
accreting at close to the Eddington limit (within a factor of a few).
Their name comes from the characteristic pattern traced out in X-ray
colour-colour diagrams [CDs] (Hasinger \& van der Klis 1989; van der
Klis 2006 and references therein). All Z sources are detected in the
radio band with approximately the same radio luminosity (Penninx 1989;
Fender \& Hendry 2000). Penninx et al. (1988) discovered a relation
between the three branches of the 'Z' in the CD and the strength of
the radio emission (see Migliari \& Fender 2006 for further discussion
and references). The two Z sources in which the radio emission has
been spatially resolved, Sco X-1 and Cir X-1, are both found to be
associated with highly relativistic flows energising slower-moving
radio emitting components (Fomalont et al. 2001; Fender et al. 2004).
In particular, the most relativistic flow yet identified in our galaxy
is associated with the neutron star jet source Cir X-1 (Fender et
al. 2004) which sometimes displays X-ray spectral and timing
characteristics similar to that of the Z sources. SS 433 may also
display this characteristic of an unseen, fast flow energising
slower-moving knots (Migliari et al. 2005), which is not something
which has been observed in any {\em bona fide} black hole candidate.

\subsection{XTE J1701-462: a new, transient, Z source}

XTE J1701-462 was discovered by the RXTE satellite as a bright new
X-ray transient on 2006 Jan 18 (Remillard et al. 2006). X-ray
observations soon indicated that it was likely to be a 'new' Z source
(Homan et al. 2006a,b). An infrared counterpart was reported (Maitra
et al. 2006; Maitra \& Bailyn 2006), which was found to be coincident
with a radio source (Fender et al. 2006), although formally
inconsistent with a localisation from the Swift X-ray Telescope
(Kennea et al. 2006). Subsequent localisation to better than one
arcsec with {\em Chandra} confirmed the association with the optical /
infrared / radio counterparts (Krauss et al. 2006).

Homan et al. (2007) recently reported the first two months' of RXTE
observations of the source, summarising the evidence for its
classification as a new Z source.  Colour-colour and
Hardness-Intensity diagrams (CD, HID respectively) are commonly used
to identify patterns of X-ray behaviour in galactic accreting X-ray
sources (e.g. Homan \& Belloni 2005; van der Klis 2006).  In Homan et
al. (2007) it was noted that the pattern XTE J1701-462 traced out in
the CD / HID changed significantly on at least nine occasions in the
first 70 days of the outburst. In particular it was noted that for the
first $\sim 25$ days of the outburst (i.e. up to around 2006 Feb 16 /
MJD 53782) the source was `Cyg-like', and subsequently most
`Sco-like', these labels referring to apparent subgroups within the Z
sources (see Homan et al. 2007 for more details).

\section{Observations}

XTE J1701-462 was observed at sixteen epochs with the Australia
Telescope Compact Array (ATCA) between 2006 Jan 22 -- April 23.  All
observations were made simultaneously at 4.8 and 8.6 GHz.  A log of
these observations is presented in table 1, and Fig 1 indicates the
observation epochs on a lightcurve of the X-ray emission. Many of the
observations were very short (nine of the sixteen were less than 1.5
hr on source) which resulted in some difficulties in mapping the
source. Six of the observations (A, B, F, K, M, N) resulted in
unambiguous detections.

No polarised signal was detected at any time, with an upper limit on
the linear and circular polarisation of $\leq 6$\% for epoch M, when
the source was brightest.

\begin{figure}
\epsfig{file=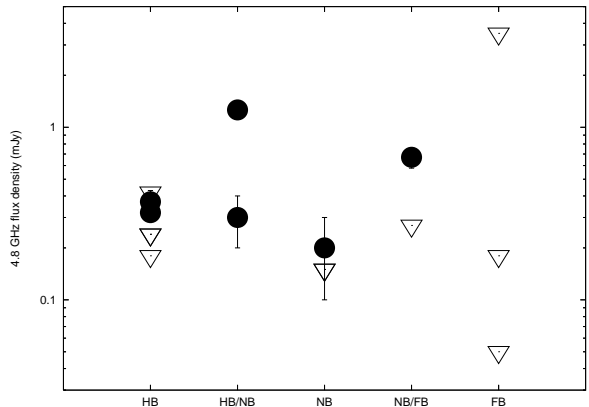, angle=270, width=8cm}
\caption{Radio detections (filled circles) and upper limits (open
triangles) as a function of our estimate of the branch / branches of
the Z the source was on at / close to the time of the radio
observations (see table 1). The results are consistent with, but do
not independently establish, the relation claimed for the Z source GX
17+2, in which radio emission is strongest on the HB/NB, and
suppressed on the FB.}
\end{figure}

Data reduction was performed with the MIRIAD package (Sault, Teuben \&
Wright 1995). In most cases naturally weighted, cleaned maps, were
used for image analysis. We were able to produce a combined map from
all the data using visibilities calibrated individually at each epoch.
Fig 2 presents a 4.8 GHz map of the field of XTE J1701-462 in which
all sixteen datasets were combined, resulting in slightly more than 24
hr on-source. The radio counterpart of XTE J1701-462 is clearly
detected at RA 17:00:58.43 Dec -46:11:08.44, with an uncertainty of
about 0.3 arcsec in each coordinate.

\begin{figure}
\epsfig{file=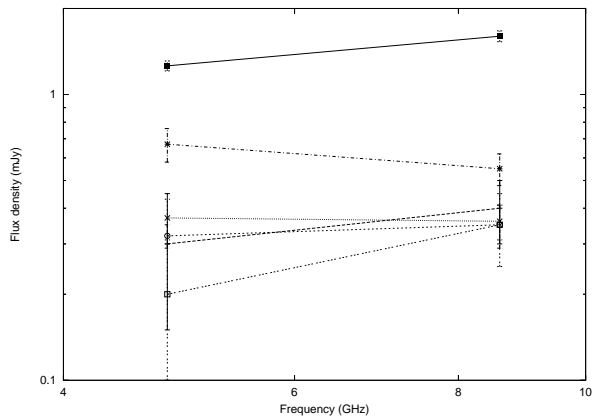, angle=270, width=8cm}
\caption{Two-point radio spectra for XTE J1701-462 for the six
unambiguous detections. The spectra are generally flat or inverted, in
contrast to the optically thin emission usually associated with
transients. If real, this indicates that when the radio emission is
detected the jet is currently being generated in regions with
significant synchrotron self-absorption (i.e. the base of the jet
close to the neutron star).}
\end{figure}

\section{Discussion}

\subsection{Coupling between radio and X-ray emission}

In order to compare the radio and X-ray properties of XTE J1701--462,
we have made an estimate of which part of the 'Z' the source was on at
or close to the time of each radio observation. XTE J1701--462 was
observed 97 times with RXTE between 2006 Jan 22 - March 20, and once
on 2006 April 23, providing simultaneous X-ray coverage for three of
our radio observations and X-ray covergage within a few hours for the
others. The RXTE observations before March 20 were all classified
according to their timing properties and position in X-ray colour-colour
diagrams by Homan et al.\ (2007) and a separate analysis of the 2006
April 23 RXTE observation was done for this paper. The location along
the Z tracks was estimated for the RXTE observations that were closest
in time to our radio observations, by determining their position
(upper/middle/lower) along the branches of full Z tracks that were
traced out in various time intervals (see Fig. 3 in Homan et al.\
2007).  These estimates are indicated in Table 1. The radio detections
and upper limits are also indicated as a function of branch of the Z
in Fig 2.

Our observations appear to indicate that XTE J1701-462 is most likely
to be detected as a radio source when on the horizontal or normal
branches (HB, NB) of the Z.  There are no detections on the flaring
branch, including the most stringent upper limit (observation P).
This is consistent with the behaviour noted first noted for GX 17+2 by
Penninx et al. (1988) and possibly universal for all Z sources
(Penninx 1989). These periods of relatively strong radio emission
almost certainly correspond to the formation of jet-like outflows
and/or their interaction with the surrounding medium (as directly
observed in the Z/Z-like sources Sco X-1 and Cir X-1; Fomalont et
al. 2001; Fender et al. 2004). In particular, based primarily on
results for Sco X-1, Migliari \& Fender (2006) suggest that the HB /
NB vertex may correspond to the point at which the most powerful eject
events occur, and that the jet is suppressed on the FB.

The Z sources are likely to be accreting persistently (in most cases)
at close to the Eddington limit. It may be useful to
compare them to black hole X-ray binaries accreting at comparably high
Eddington ratios, such as GRS 1915+105. Both classes of object make
dramatic and rapid state transition and are associated with episodic
production of powerful relativistic jets. Such objects may, in turn,
be our best `local' equivalents of quasars accreting at very high
rates at redshifts of $z \geq 1$. Therefore careful comparison of
objects such as XTE J1701-462 with black holes accreting at high rates
may provide our best test of the effects of e.g. event horizons,
static limits on accretion and jet formation.

\subsection{Spectral index}

The two-point radio spectra of these detections are plotted in Fig
3. It is interesting to note that for the majority of the radio
detections of XTE J1701-462 the source radio spectrum is flat /
inverted (spectral index $\alpha \geq 0$, where S$_{\nu} \propto
\nu^{\alpha}$). This seems to be in contrast to most emission
associated with transient outbursts from X-ray binaries which
generally has an optically thin spectrum ($\alpha \leq
-0.6$). Optically thin emission is also observed in Cir X-1 (Fender et
al. 1998; 2004) and Sco X-1 (Fomalont et al. 2001), although in both
sources episodes of flat-spectrum core emission has been seen. The
flat/inverted radio spectrum is in fact more reminiscent of the
steady, flat-spectrum radio emission observed from black holes in hard
X-ray states (Fender 2001), although Migliari \& Fender (2006) do
suggest that it should also be observed in hard state neutron
stars. Such flat-spectrum emission is believed to arise in a partially
self-absorbed jet (Blandford \& K\"onigl 1979; Kaiser 2006). Given the
rather poor coverage and weak source in most cases, there remains some
uncertainty in these spectral index measurements.  If the
spectrum is genuinely flat or inverted it implies the ongoing
production of a compact jet and not, for example, shocks in diffuse
regions well separated from the binary. The limits on the polarisation
($\leq 6$\% in Stokes Q, U, V) are consistent with the self-absorbed
jet model -- measured linear polarisations for flat-spectrum jets in
black hole X-ray binaries are at the few \% level (Fender 2001 and
references therein).

\subsection{Luminosity and distance}

\begin{figure}
\epsfig{file=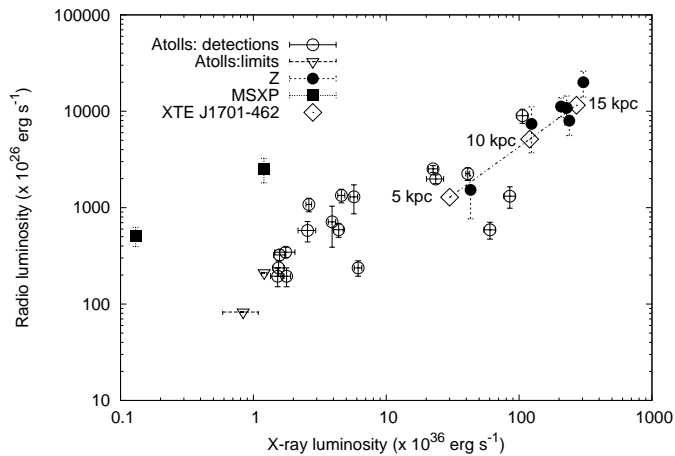, angle=270, width=9cm}
\caption{Radio luminosity as a function of X-ray luminosity for
neutron star X-ray binaries. The Z sources are the most luminous group
of objects in both bands. The locus of XTE J1701-462, around the time
of our observation M, is indicated for three different distances.}
\label{lum}
\end{figure}

As noted above, it is a defining characteristic of Z sources that they
are accreting at near to (sometimes in excess of) the Eddington limit
for a 1.4 solar mass neutron star, and that they are relatively strong
radio sources. Migliari \& Fender (2006) plotted the radio luminosity
as a function of X-ray luminosity for all neutron star X-ray binaries
with radio counterparts, and indeed found the Z sources to be the most
luminous in both bands. In Fig {\ref{lum}} we plot the same sample
with a point representing XTE J1701-462 at each of three distances: 5,
10 and 15 kpc. The point is based upon the estimated flux of $\sim
10^{-8}$ erg s$^{-1}$ cm$^{-2}$ for observation phase G in Homan et
al. (2007), which is close in time to our observation M. We have
chosen a mean flux density for this period of $\sim 0.5$ mJy. We note
that while this estimate of the mean radio flux does not take into
account non-detections, nor did the estimates of Fender \& Hendry
(2000) on which the points for the other Z sources are based. Based
solely on this rough comparison, it seems reasonable to conclude that
XTE J1701-462 lies at a distance of $\sim 10$ kpc, with an error of a
factor of two. Compare this to a different approach in Homan et
al. (2007), where a distance of $\sim 15$ kpc was derived.

\subsection{An ultrarelativistic jet and a large-scale nebula ?}

Finally, the similarities with the other Z sources, in terms of the
X-ray properties and their relation to the radio emission, suggest
that XTE J1701-462 may harbour a relativistic jet like those resolved
in Sco X-1 (Fomalont et al. 2001) and Cir X-1 (Fender et
al. 2004). Future high angular observations with ATCA and Australian
e-VLBI (Philips et al. 2007) may resolve and track the variability of
such a jet.

\begin{figure}
\epsfig{file=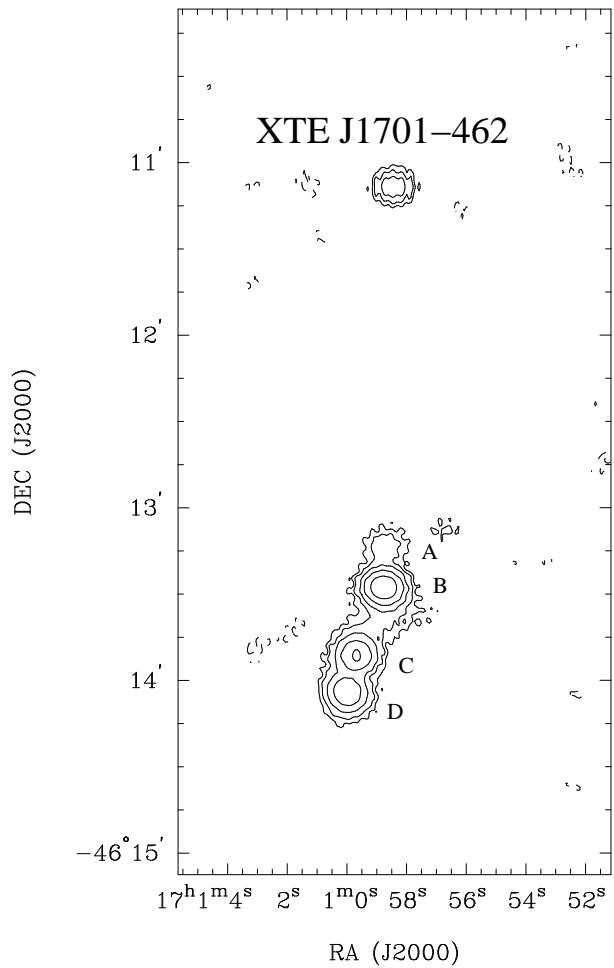, angle=0, width=8cm}
\caption{Summed radio image from all sixteen epochs of ATCA
observations (24.1 hr on source). The radio counterpart of XTE
J1701-462, unambiguously detected in at least one third of the
observations, is indicated to the north of the figure. This radio
source is entirely consistent with the {\em Chandra} localisation
reported by Krauss et al. (2006). There is clearly a complex of radio
sources about three arcmin to the south, and aligned towards, the
X-ray binary. Peak fluxes for XTE J1701-462 and components A--B
from this map are given in Table 2. The image is at 4.8 GHz and the
contours are at (-3,3,6,12,24) times the r.m.s. noise of 28$\mu$Jy.}
\label{wide}
\end{figure}

Furthermore, Cir X-1 has an arcmin-scale radio nebula which seems to
be powered by the central jet (Stewart et al. 1991; Tudose et
al. 2006), as do a number of black hole X-ray binaries (e.g. Mirabel
et al. 1992; Corbel et al. 2002). In order to investigate whether XTE
J1701-462 may have a similar large-scale radio structure, we created a
deep image by summing all of the available data
(Fig\ref{wide}). Several fairly strong sources are clearly detected to
the south of XTE J1701-462, and are approximately lined up back
towards the X-ray binary. Peak fluxes for XTE J1701-462 and the
sources A--D are listed in Table 2, measured from naturally weighted
summed maps at both frequencies. The spectral index is also given, and
indicates that sources A--D are all nonthermal in nature, with an
apparent trend of steepening spectra with increasing distance to the
south (although resolution effects may play a role if the structure
also becomes more diffuse to the south).

\begin{table}
\caption{Peak fluxes and spectral index of XTE J1701-462 and sources
A--D at 4.8 and 8.6 GHz as measured from the summed, naturally weighted maps.}
\begin{tabular}{cccc}
Source & $F_{\rm 4.8}$ (mJy/beam) & $F_{\rm 8.6}$ (mJy/beam) & $\alpha_{4.8-8.6}$\\
XTE J1701-462 & 0.6 & 0.7 & +0.3\\
A & 0.3 & 0.2 & -0.7 \\
B & 2.5 & 0.9 & -1.7\\
C & 2.0 & 0.5 & -2.4 \\
D & 1.5 & 0.3 & -2.7 \\
\end{tabular}
\end{table}

Note that while variations in the flux of XTE J1701-462 between epochs
could cause artefacts in such a combined map, we consider it highly
unlikely that they could produce the extended structure seen in
Fig\ref{wide}. Furthermore, this structure is visible to a varying
degree in longer individual runs (runs B, F, M, N, O and P). The
obvious initial interpretation of this structure is that it is an
unrelated background radio source, probably extragalactic. However,
based on the source counts from the ATESP 5 GHz deep survey (Prandoni
et al. 2006), we would only expect $\sim 0.05$ sources in the 1.5-2.5
mJy range in a region the size of the map presented in
Fig\ref{wide}. Sources B, C and D are all in this range.  In addition,
if we consider that sources A--D are a single extended source, for the
integrated flux density, $\sim 5$ mJy, the ATESP survery indicates a
mean angular size of $\la 10$ arcsec, much less than the $\ga 60$
arcsec angular extent of the source(s). Furthermore, the SIMBAD
{\footnote{\bf simbad.u-strasbg.fr}} database lists no catalogued
objects at this location. XTE J1701-462 appears to lie in a region of
the sky which is overdense in bright, compact, 5 GHz radio sources.

The possibility must therefore be considered that the source to the
south may be related in some way to the X-ray binary. If so, it
appears to take the form of a curved, one-sided radio jet (there is no
comparable structure to the north). At an angular separation from the
X-ray binary of three arcmin and a distance of 5-15 kpc, the physical
separation of components A--D from XTE J1701-462 greater than
$10^{19}$ cm, which at face value seems to rule out an association
with the 2006 January outburst, as apparent velocities greater than
$80c$ would be required (but see Fender et al. 2004 for the
intriguingly similar case of Cir X-1).  Further radio observations to look for
variability and more diffuse structure are planned, as are
observations at other wavelengths.

\section{Conclusions}

In 2006 Janauary a new X-ray transient, XTE J1701-462, was discovered,
and rapidly established to display the characteristics of a Z-type
neutron star X-ray binary. In this paper we report on the variable
radio counterpart to this X-ray source, which demonstrates a coupling
between X-ray state (branch of the Z) and radio luminosity similar to
that of the other Z sources. By analogy with Sco X-1 and Cir X-1, we
interpret this as telling us about the connection between accretion
flows and the production of a relativistic jet. Fortunately, since its
activation, XTE J1701-462 has remained a bright X-ray source (as of
2007 May) and provides us with a new laboratory in which to study
these phemonena. Furthermore, there is some evidence that the source
is powering a large-scale radio jet, which hints not only at phases of
past activity but also at the possibility of a nebula which may be
used to constrain the integrated jet power and provide a further
direct comparison between jets from neutron stars and those from black
holes.

\section{Acknowledgements}

RPF would like to thank Simone Migliari for providing his data for Fig
4, and Derek Moss for assistance with radio source counts. We would
also like to thank the referee for suggesting, and the editor for
insisting, that we check on the expected source counts. We would like
to acknowledge the extreme flexibility of the schedulers and observers
at the Australia Telescope Compact Array who supported these
observations. The ATCA is run by ATNF for CSIRO.

\end{document}